# Nonadiabatic Induced Dipole Moment by High Intensity Femtosecond Optical Pulses

**I. G. Koprinkov**

*Department of Applied Physics, Technical University of Sofia, 1756 Sofia, Bulgaria*
*Fax: (+359 2) 683215; E-mail: igk@tu-sofia.bg*

**Abstract**: Nonadiabtic dressed states and nonadiabatic induced dipole moment in the leading order of nonadiabaticity is proposed. The nonadiabatic induced dipole moment is studied in the femtosecond time domain.

### 1. Introduction

The interaction of high-intensity ultrashort optical pulses with atomic systems is attracting increasing attention in parallel with the advances in the high-intensity femtosecond laser technology. Some of the widely used approaches in the field-matter interaction problem are not as valid at extreme conditions of such interactions. The *conventional* adiabatic approximation, *i.e.*, neglecting the field time derivatives [1], is among these. For femtosecond pulses, the field time derivatives are not already small and their contribution cannot be neglected. Evidence of nonadiabatic effects were found experimentally and confirmed numerically (solving the time dependent Schrödinger equation) for sub-*100 fs* pulses in the form of increased order and intensity of the generated high-harmonics [2]. The numerical simulations within the strong field approximation predict that the adiabatic approximation breaks for, *e.g.*, sub-*27 fs* pulses in the case of argon atom at $3x10^{14} W/cm^2$ peak intensity [3]. Nonadiabatic dressed states (NADSs) have been recently derived [4] introducing a *generalized* adiabatic condition (GAC) [4, 5]. The NADSs include nonadiabatic terms (phase and amplitude time derivatives of the field) up to given order, which still allow solving the quantum equations of motion, keeping at the same time the natural generalization of the corresponding quantities from the adiabatic to the nonadiabatic case.

Here we present NADSs of leading order of nonadiabaticity [6]. This simplifies the analytic study of the nonadiabatic effects and makes their interpretation more easy and tractable. An analytic expression for the induced dipole moment within such NADSs is found and studied in the femtosecond time domain. Ways to extend the application of the NADSs toward the ultrashort time scale are considered in this work.

### 2. Nonadiabatic dressed states and nonadiabatic induced dipole moment

The GAC imposes the requirement that the field time derivatives $\partial_t^n(\partial_t\varphi - i\Omega^{-1}\partial_t\Omega)$ must be much smaller than the *product* of the respective powers of frequency detuning $\Delta\omega$ from the atomic resonance (Fig.1) and the Rabi frequency $\Omega = \mu E/\hbar$. Thus, the range of validity of the NADSs can be extended at large frequency detuning, and/or high field amplitude $E$. Fortunately, these are common conditions for the high field physics. If the frequency detuning is large enough (such a case is taken as an example here, see below), the GAC can be satisfied even if the nonadiabatic terms are not much smaller but comparable to, or even exceeding to some extent, the Rabi frequency. The shortcoming of such an approach is that the rotating wave approximation (RWA), also used in the derivation of the NADSs, will be violated at too large detuning. In that case, the absolute value of the predicted results must be taken with caution, but the relative one (versus the corresponding adiabatic case) is expected to be a good indication of nonadiabaticity as the RWA is used in both, the adiabatic [1] and the nonadiabatic theory [4]. Another approach, increasing the field amplitude, can also be exploited because the NADSs represent a nonperturbative solution with respect to the field strength. In this way, for given parameter range, the NADSs can be extended approaching the attosecond time scale, subject also to the validity of the envelope-carrier concept for the optical field. The later (which does not directly concern the NADSs solution itself) allows expression of the field in a convenient and intuitively clear form. As has been found, the envelope-carrier concept of the optical field is "legitimate for FWHM pulse durations down to the carrier oscillation cycle" [7].

The ground $|G\rangle$ and the excited $|E\rangle$ NADSs for a two level system are expressed in the form [4]

$$
\begin{aligned}
|E\rangle &= COS(\theta/2)|E\rangle_r - SIN(\theta/2)|E\rangle_v \\
|G\rangle &= SIN(\theta/2)|G\rangle_v + COS(\theta/2)|G\rangle_r
\end{aligned} \quad , \tag{1}
$$

where the real (index "r") and virtual (index "v") components of the NADSs, Fig.1, are

$$
\begin{aligned}
|G\rangle_r &= |1\rangle \exp\left\{-i\int_0^t \omega_G dt'\right\} & |E\rangle_r &= |2\rangle \exp\left\{-i\left[\int_0^t \widetilde{\omega}'_E dt' + \varphi(t)\right]\right\} \\
|G\rangle_v &= |2\rangle \exp\left\{-i\left[\int_0^t (\omega_G + \omega)dt' + \varphi(t)\right]\right\} & |E\rangle_v &= |1\rangle \exp\left\{-i\int_0^t (\widetilde{\omega}'_E - \omega)dt'\right\}
\end{aligned} \tag{2}
$$

The complex "weight" factors $COS(\theta/2) = (\widetilde{\Lambda}'_1 / \widetilde{\Omega}')^{1/2}$ and $SIN(\theta/2) = (-\widetilde{\Lambda}'_2 / \widetilde{\Omega}')^{1/2}$ determine the partial presentation of the real and virtual components, respectively, in the NADSs, and $\widetilde{\Omega}'$ is the instantaneous off-resonant Rabi frequency.

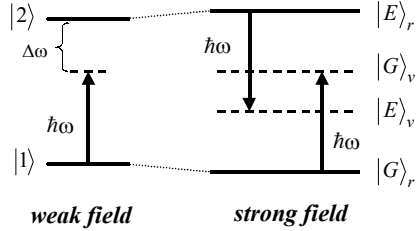

Fig.1. The bare (weak field) and the dressed (strong field) states of a two-level quantum system.

The results presented below are derived using the following additional conditions: *i*) only the leading order nonadiabatic term $\Omega^{-1}\partial_t \Omega$ is retained in the NADSs *ii*) the dumping is neglected, *iii*) chirp free laser pulses are considered. The parameters of the NADSs then become:

$$\widetilde{\Omega}' = [\Delta\omega^2 + \Omega^2 + i2\Delta\omega \, \Omega^{-1}\partial_t\Omega]^{1/2} \tag{3}$$

$$\widetilde{\Lambda}'_1 = \frac{1}{2}[\Delta\omega + i\,\Omega^{-1}\partial_t\Omega + (\Delta\omega^2 + \Omega^2 + i2\Delta\omega \, \Omega^{-1}\partial_t\Omega)^{1/2}] \tag{4}$$

$$\widetilde{\Lambda}'_2 = \frac{1}{2}[\Delta\omega + i\,\Omega^{-1}\partial_t\Omega - (\Delta\omega^2 + \Omega^2 + i2\Delta\omega \, \Omega^{-1}\partial_t\Omega)^{1/2}] \tag{5}$$

Equations (1)-(5) allow finding of the nonadiabatic induced dipole moment in the corresponding NADSs, $\mu_{nad} = \langle G|\hat{\mu}|G\rangle$ and $\mu_{nad} = \langle E|\hat{\mu}|E\rangle$. Two terms contribute to $\mu_{nad} = \langle G|\hat{\mu}|G\rangle$, an *in-phase* term at the driving field,

$$\mu^{ip}_{nad} = \mu\sqrt{\left[\sqrt{(D-M)^2 - N^2} + \left(D - \sqrt{M^2 + N^2}\right)\right]}(2D)^{-1}$$

and (much smaller) out-of-phase term,

$$\mu_{nad}^{out} = \mu \sqrt{\left[\sqrt{(D-M)^2 - N^2} - \left(D - \sqrt{M^2 + N^2}\right)\right](2D)^{-1}} \quad ,$$

where $D = (\Delta\omega^2 + \Omega^2)^2 + 4\Delta\omega^2(\Omega^{-1}\partial_t\Omega)^2$, $M = \Delta\omega^2(\Delta\omega^2 + \Omega^2) + (\Omega^{-1}\partial_t\Omega)^2(3\Delta\Omega^2 - \Omega^2)$, $N = 2\Delta\omega \ (\Omega^{-1}\partial_t\Omega)[\Omega^2 + (\Omega^{-1}\partial_t\Omega)^2]$, and $\mu \equiv \langle 2|\hat{\mu}|1\rangle = \langle 1|\hat{\mu}|2\rangle$ is the dipole matrix in the bare states. If the condition $\Delta\omega \gg \Omega,\ \Omega^{-1}\partial_t\Omega$ (a more relaxed one than the GAC) is satisfied, then $\mu_{nad}^{ip}$ takes much simpler form

$$\mu_{nad}^{ip} \approx \mu \sqrt{(\Delta\omega^2 + \Omega^2)\,[\Omega^2 + (\Omega^{-1}\partial_t\Omega)^2]\,[(\Delta\omega^2 + \Omega^2)^2 + 4\Delta\omega^2(\Omega^{-1}\partial_t\Omega)^2]^{-1}} \quad , \tag{6}$$

where $\mu_{nad}^{ip}$ represents a natural generalization of the induced dipole moment in the adiabatic dressed states, $\mu_{ad} \equiv \langle G|\hat{\mu}|G\rangle = \mu\,\Omega\,(\Delta\omega^2 + \Omega^2)^{-1/2}$ [1]. The later can be reproduced from $\mu_{nad}^{ip}$, Eq. (6), ignoring the nonadiabatic term $\Omega^{-1}\partial_t\Omega$. The influence of the laser phase/chirp on the NADSs will be considered in a forthcoming work.

### 3. Results and discussion

The nonadiabatic induced dipole moment $\mu_{nad}^{ip}$, Eq.(6), versus time duration of the driving pulse has been studied for two different values of the dipole moment/Rabi frequency, a moderate ($\mu \approx 0.01D$) and a high ($\mu \approx 1D$) one. The medium is assumed highly transparent, $\Delta\omega \approx 10^{16}\ s^{-1}$. The results for a chirp free $sech^2(t/\tau)$ pulse of $5\times10^{13}$ W/cm$^2$ peak intensity are shown in the Fig.2 taking into account the range of validity of the GAC. At the specified parameters, the most critical condition (within the GAC) in the derivation of the expression for $\mu_{nad}^{ip}$ is $\tau^{-3} \ll \Omega^2\Delta\omega$, where $\tau$ is the pulse duration.

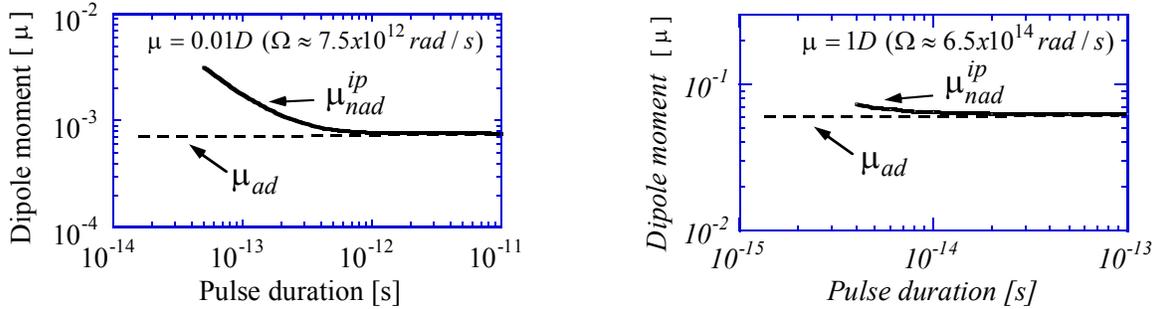

Fig.2. The nonadiabatic induced dipole moment versus pulse duration for a two-level system at moderate $\mu \approx 0.01D$ (a), and high $\mu \approx 1D$ (b) values of the dipole moment.

The results show substantial increasing of the magnitude of the induced dipole moment (relatively its adiabatic limit $\mu_{ad}$) with shortening of the driving pulse if one accounts for the time derivative of the field amplitude. It is much stronger expressed for low value of the Rabi frequency. Such an increase of the induced dipole moment, together with the increased dipole acceleration, can be related with the experimentally observed increasing of, not only the order, but also the intensity of the generated high harmonics at shorter laser pulse [2]. $\mu_{nad}^{oop}$ grows much more rapidly than $\mu_{nad}^{ip}$ with shortening of the laser pulse, while remaining much smaller in the range under considerations, Fig.3. It is important to note that the nonadiabatic factor $\Omega^{-1}\partial_t\Omega$ may not simply become large but it may exceed the Rabi frequency for a range of realistic conditions in the femtosecond time domain. Thus, for example, the Rabi frequency at $10^{14}$W/cm$^2$ is $\Omega \approx 8.7\times10^{12}s^{-1}$ (at $\mu \approx 0.01D$) while $\Omega^{-1}\partial_t\Omega$ for 100fs sech$^2$ pulse reaches values of about $1.5\times10^{13}s^{-1}$. In this way, a number of strong field phenomena in the femtosecond time domain become substantially nonadiabatic. This, in particular, means that the simplified treatment of high harmonic generation in terms of an adiabatic two-level theory [8] should be more realistic within the nonadiabatic approach.

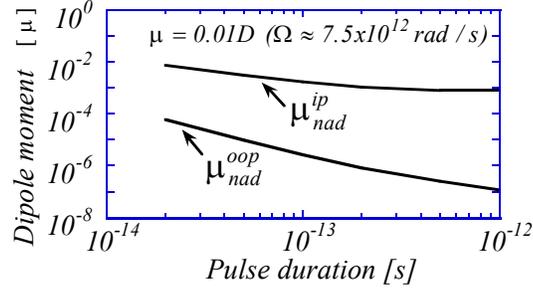

Fig. 3. In-phase and out-of-phase nonadiabatic dipole moment versus pulse duration.

The present theory is inherently nonadiabatic (in the time domain), as the nonadiabatic factors (net variation of the field amplitude E in the normalized Rabi frequency time derivative, $\Omega^{-1}\partial_t\Omega$) are explicitly included in the solution of the quantum states. This means that the nonadiabatic effects will be involved in all considerations based on the NADSs. A basic appearance of the nonadiabaticity is the transition between different adiabatic states. Another effect of nonadiabaticity is the appearance of time lag between the ultrashort driving field and the induced dipole moment [9]. With the NADSs we introduce a new aspect of nonadiabticity, *i.e.*, the explicit contribution of the field time derivatives (neglected in the adiabatic approach [1]) to the field-matter interaction. This reveals the *non-instantaneousness* of the interaction, which becomes "visible" at ultrashort pulse excitation. In this work we focus on the *pure radiative* part of the interaction of a quantum system with a nonadiabatic field in terms of the induced dipole moment within given dressed state. This differs from the conventional (*e.g.*, Landau-Zener-Stueckelberg [10]) theory of nonadiabatic transitions, which emphasis on the transition between adiabatic states. The later is more relevant to the nonadiabaticity in *space domain*, even if the adiabatic parameter depends explicitly on time, as it usually implies nonadiabatic transitions between electronic states by gaining energy from fast (nonadiabatic) nuclear motion. The nonadiabatic factors play role of coupling (*e.g.*, the nuclear momentum operator) that causes nonadiabatic transition between, otherwise, adiabatic states. Transition between different NADSs, in our case $|G\rangle$ and $|E\rangle$, also takes place but it is beyond the scope of the present work.

Causal implication of the material phase along with the field nonadiabatic factors, called *material phase tracking*, has been found for the first time in [5]. Existence of observable physical consequences of the material phase is supported due to evidences from theoretical and experimental point of view [5, 11]. A marked appearance of the material phase in the physical processes can be also found in some high field phenomena. The strong field approach to high harmonic generation shows that the physical action $S$ (*i.e.*, the material phase $\Phi = -\hbar^{-1}S$) acquired by the electron along a given trajectory has observable physical consequences as it influences some observable features of the generated harmonics [12].

**4. Conclusions**

Simplified and tractable form of the nonadiabatic dressed states is presented and the way to extend their validity for the high field ultrashort optical phenomena is discussed. A closed form expression of the induced nonadiabatic dipole moment was found. The optical phenomena in the femtosecond time domain were found to become substantially nonadiabatic, subject also to the field strength/Rabi frequency. To the best of our knowledge, this is the first explicit analytic treatment of optical nonadiabatic effects in the femtosecond region, closely approaching the attosecond time scale.